\documentclass[12pt,notitlepage]{article}

\usepackage{amsmath,amssymb}
\usepackage{graphicx}

\usepackage{color}

\def\ignore#1{{}}

\let\oldtheequation=\theequation
\def\doteqs#1{\setcounter{equation}{0}            
	\def\theequation{{#1}.\oldtheequation}}
\newcounter{sxn}
\def\sx#1{\addtocounter{sxn}{1} \vskip 1.cm  \goodbreak
	\noindent{\large\bf\leftline{\thesxn.~~#1}} \nobreak \vskip -.5cm}
\def\sxn#1{\sx{#1} \doteqs{\thesxn}}

\newcounter{axn}

\newdimen\mybaselineskip
\newcommand{\beeq}{\begin{equation}}
	\newcommand{\eneq}{\end{equation}}
\newcommand{\beqn}{\begin{eqnarray}}
	\newcommand{\eeqn}{\end{eqnarray}}

\newcommand{\ba}{\begin{array}}
	\newcommand{\ea}{\end{array}}

\newcommand{\be}{\begin{equation}}
	\newcommand{\ee}{\end{equation}}
\newcommand{\bea}{\begin{eqnarray}}
	\newcommand{\eea}{\end{eqnarray}}
\newcommand{\beal}{\setcounter{letter}{1} \begin{eqnarray}}
	\newcommand{\eeal}{\addtocounter{equation}{1} \end{eqnarray}}

\newcommand{\larrow}{\,\,\,\,\hbox to 30pt{\rightarrowfill}
	\,\,\,\,}
\newcommand{\slarrow}{\,\,\,\hbox to 20pt{\rightarrowfill}
	\,\,\,}

\def\la{\raise.16ex\hbox{$\langle$}\lower.16ex\hbox{}  }
\def\ra{\, \raise.16ex\hbox{$\rangle$}\lower.16ex\hbox{} }

\def\psibar{ \psi \kern-.65em\raise.6em\hbox{$-$} \lower.6em\hbox{} }
\def\psibarb{ \psi \kern-.65em\raise.6em\hbox{$-$}  }

\begin{document}

\thispagestyle{empty}

\baselineskip=12pt



\vspace*{3.cm}

\begin{center}  
{\LARGE \bf  Calculating quasinormal modes of Schwarzschild anti-de Sitter black holes using the continued fraction method}
\end{center}

\baselineskip=14pt

\vspace{2cm}
\begin{center}
{\bf  Ramin G. Daghigh$^1$, Michael D. Green$^2$, and  Jodin C. Morey$^3$}
\end{center}

\vspace{0.25 cm}
\centerline{\small \it $^1$ Natural Sciences Department, Metropolitan State University, Saint Paul, Minnesota, USA 55106}
\vskip 0 cm
\centerline{} 

\centerline{\small \it $^2$ Mathematics and Statistics Department, Metropolitan State University, Saint Paul, Minnesota, USA 55106}
\vskip 0 cm
\centerline{} 

\centerline{\small \it $^3$ School of Mathematics, University of Minnesota, Minneapolis, Minnesota, USA 55455}
\vskip 0 cm
\centerline{}

\vspace{1cm}
\begin{abstract}
We investigate the scalar, gravitational, and electromagnetic quasinormal mode spectra of Schwarzschild anti-de Sitter black holes using the numerical continued fraction method.  The spectra have similar, almost linear structures.  With a few exceptions, the low overtone quasinormal modes are consistent with previously obtained results in the literature that use other numerical techniques.   The intermediate and high overtone quasinormal modes, in comparison to the Schwarzschild case, converge very quickly to the asymptotic formulas previously obtained by analytic monodromy techniques.  
In addition, we find a connection between the analytic asymptotic formulas and the purely imaginary modes.  In particular, these formulas can be used to predict the bifurcation of the lowest damped electromagnetic modes.  Finally, we find no high overtone quasinormal modes with high oscillation frequency and low damping, which had been previously predicted.


\baselineskip=20pt plus 1pt minus 1pt
\end{abstract}

\newpage

\sxn{Introduction}

\vskip 0.3cm

 Black holes in anti-de Sitter (AdS) spacetime have attracted a great deal of attention for multiple reasons.  It was shown by Hawking and Page \cite{Page} that, unlike black holes in flat spacetime, large black holes in AdS spacetime have positive specific heat and can be in stable equilibrium with thermal radiation at a fixed temperature.  Also, according to the AdS/CFT correspondence proposed by Maldacena \cite{Maldacena}, a large static black hole in AdS spacetime corresponds to a thermal state of a system in conformal field theory (CFT).   
 More specifically, the dynamical time scale for the return to thermal equilibrium in CFT is equal to the decay rate of the perturbation of a large black hole in AdS spacetime.

It is, therefore, important to determine the natural vibrational modes of perturbations for black holes in AdS spacetime.  These vibrational modes are called quasinormal modes (QNMs), which are discrete and complex. The imaginary part of the frequency signals the presence of damping, a necessary consequence of boundary conditions that require energy to be carried away from the system.

The significance and stability of QNMs, when discontinuities are introduced to the Regge-Wheeler or QNM potential, were investigated initially by Nollert in \cite{Nollert1}.  Later, we refined and expanded upon Nollert's results in \cite{DGM-significance}.  This investigation was followed by multiple papers \cite{QNM-Significance1, QNM-Significance2, QNM-Significance3, QNM-Significance4, QNM-Significance5, QNM-Significance6} on the stability of the QNM spectrum.  It has been shown that QNMs (including the least damped mode) are unstable when discontinuities are introduced to the Regge-Wheeler potential.  However, the ringdown waveform stays stable in the presence of small discontinuities.  Therefore, one can still use the QNM spectrum of the smooth potential to describe the behavior of the ringdown waveform.  Consequently, the stability issue is not a deterrent for us in this paper.  In addition, since our universe is not an AdS space, the importance of AdS lies mainly in the theoretical and mathematical realm where we do not need to introduce physical discontinuities (such as the presence of matter in the vicinity of a black hole) to the QNM potential.  

The ringdown waveform of Schwarzschild-AdS black holes for scalar perturbations was investigated by Chan and Mann in \cite{ChanMann} where some of the lowest damped modes in certain cases were calculated by finding the roots of the frequency domain Green's function.
Later, Horowitz and Hubeny \cite{Horowitz} calculated the low overtone QNMs of Schwarzschild-AdS black holes for scalar perturbations in four, five, and seven spacetime dimensions, which are of interest in the context of the AdS/CFT correspondence.  This work was completed by Zhu {\it et al.} \cite{Zhu-smallBH} and Konoplya \cite{Konoplya0}, who analyzed the scalar QNMs for small Schwarzschild-AdS black holes in detail.  Cardoso and Lemos \cite{Cardoso-L} extended these calculations to include electromagnetic and gravitational perturbations.  In addition, Moss and Norman \cite{Moss} calculated roughly twenty QNM roots for the gravitational perturbations of the Schwarzschild-AdS black hole and showed the duality between axial and polar perturbations can be preserved if we apply Dirichlet boundary conditions to axial perturbations, and Robin boundary conditions to polar perturbations.  Berti and Kokkotas confirmed all the above results in \cite{Berti-K} and extended their numerical calculations to Reissner-Nordstr\"om black holes in AdS spacetime.  Finally, the low overtone QNMs of Dirac spinors are addressed by Giammatteo and Jing in \cite{low-tone-2}.   Motivated by AdS/CFT correspondence, many authors have extended the QNM calculations to higher spacetime dimensions (some examples are given in \cite{high-tone-5d}).  All these authors used either the method developed by Horowitz and Hubeny \cite{Horowitz} or some variation of the Fr\"obenius method.  As far as we know, no one has applied Leaver's continued fraction method \cite{Leaver} to calculate the QNMs of Schwarzschild-AdS black holes.  The continued fraction method has proven to be the most reliable numerical technique to calculate these modes.  Also, the continued fraction method with Nollert's improvement \cite{Nollert} is able to extend the numerical calculations to high overtone modes with very good precision. For a comprehensive review of the QNMs of black holes, see \cite{BertiReview} and \cite{KonoplyaReview}.

The high overtone QNMs of Schwarzschild-AdS black holes were calculated numerically for scalar, electromagnetic, and gravitational perturbations in four spacetime dimensions by Cardoso, Konoplya, and Lemos in \cite{Cardoso-K-L}.   Analytic calculations of the infinitely high overtone (asymptotic) QNMs of four dimensional Schwarzschild-AdS black holes were done by Cardoso {\it et al.} in \cite{Cardoso-N-S} using a method based on the monodromy technique developed by Motl and Neitzke \cite{Motl1, Motl2}.  Natario and Schiappa, in \cite{Natario-S}, generalized the analytic results to include Schwarzschild-AdS black holes in dimensions greater than four and Ghosh {\it et al.} \cite{Ghosh} generalized the analytic results further by studying the infinitely high overtone QNMs of asymptotically non-flat black holes in a generic way.   

Two of us in \cite{DG-HReal} adapted the analytic technique based on the complex coordinate WKB method, developed by Andersson and Howls in \cite{Andersson}, to calculate the high overtone QNMs of Schwarzschild-AdS black holes in dimensions greater than three.  In addition to confirming the previously obtained results by Natario and Schiappa, for large black holes in certain spacetime dimensions, we showed that the analytic monodromy technique implies the existence of an asymptotic region of the QNM spectrum where the real part of the QNM frequencies approaches infinity while the damping rate approaches a finite value.  The interesting thing about these ``highly real" modes is that they have a damping rate less than the least damped QNM found by Horowitz and Hubeny in \cite{Horowitz}.  Therefore, assuming they exist, these modes will be the dominant modes in the context of the AdS/CFT correspondence.  It was also shown by one of us in \cite{D-HReal} that the same highly real modes can be obtained using the monodromy technique 
developed by Natario and Schiappa in \cite{Natario-S}. The authors of \cite{Morgan} searched for these highly real modes using two different numerical methods, power series 
and time evolution, but they were not able to find them.  Since the continued fraction method with Nollert's improvement has proven to be the most effective in searching for high overtone QNMs, we use it here to search for these modes.

The paper is organized as follows.  In section 2, we lay down the general formalism.  In section 3, we describe our numerical process. In section 4, we present the results.  In section 5, we compare the numerical results with previously found analytic asymptotic formulas.  Finally, we summarize our findings in section 6.

\sxn{General Formalism}
\vskip 0.3cm
For a spherically symmetric black hole in four spacetime dimensions, all perturbations are governed generically by a Schr\"{o}dinger wave-like equation of the form
\beeq
{d^2\psi \over dr_*^2}+\left[ \omega^2-V(r) \right]\psi =0 ~,
\label{eqSchrodinger}
\eneq
where $r$ is the radial coordinate, $V(r)$ is the QNM potential and $\omega$ turns out to be the QNM frequency.  
The Tortoise coordinate $r_*$ is defined by 
\beeq
dr_* ={dr \over f(r) }~,
\label{tortoise}
\eneq
where $f(r)$ is the metric function of the form 
\beeq
f(r)=1-{2M \over r}+ \frac{r^2}{R^2}~.
\label{function f}
\eneq
Here, $M$ is the ADM mass of the black hole and $R$ is the AdS radius.\footnote{The value of the cosmological constant, $\Lambda$, is given by 
	$\Lambda =  - 3/R^2.$}
We use the geometrized unit system, where $G = c = 1$.
In this paper, we assume the perturbations depend on time as $e^{-i\omega t}$.  Consequently, in order to have damping, the imaginary part of $\omega$ must be negative.  

The effective potential in Eq.\ (\ref{eqSchrodinger}) is given in \cite{Cardoso-L, Cardoso-K-L} as
\beeq
V(r) = f(r)\left[ \frac{\ell (\ell + 1)}{r^2} - \frac{2M (s^2-1)}{r^3} + \frac{\xi}{R^2} \right],
\label{Vr}
\eneq
where $s$ is the spin of the perturbation, $\ell$ is the multipole number, and the index $\xi$ is $2$ for scalar and $0$ for electromagnetic and gravitational perturbations.  For gravitational perturbations, we only consider those with odd parity (axial perturbations).  Moss and Norman \cite{Moss} have shown that there exists a symmetry that relates odd and even parity perturbations as long as, at the AdS boundary, one chooses the conditions to be Dirichlet for odd parity and Robin for even parity perturbations.  In fact, by examining the behavior of the perturbations of the metric at the AdS boundary, it has been shown in \cite{Pufu} that
Robin boundary conditions are indeed  the appropriate ones to use for even parity perturbations.  Since we are only considering odd parity perturbations, we use Dirichlet boundary conditions.  For further discussion on issues related to different AdS boundary conditions, see for example \cite{Morgan}.

The effective potential is zero at the event horizon ($r \rightarrow r_+$).  The boundary conditions are taken so that the asymptotic behavior of the solutions is
\beeq
\psi(r) \approx \left\{ \begin{array}{ll}
                   e^{-i\omega r_*}  & \mbox{as $r \rightarrow r_+$ },\\
                   0  & \mbox{as $r \rightarrow \infty$,}
                   \end{array}
           \right.        
\label{asymptotic}
\eneq 
which represents an in-going wave at the event horizon and no waves at infinity.

The wave equation (\ref{eqSchrodinger}) in the radial coordinate is
\beeq
f{d^2\psi \over dr^2}+f'{d\psi \over dr}-\left[ \frac{\rho^2}{f}+\frac{\ell (\ell + 1)}{r^2} - \frac{2M (s^2-1)}{r^3} + \frac{\xi}{R^2} \right]\psi =0 ~,
\label{eqWaveradial}
\eneq
where $\rho=-i\omega$.
It is easy to show that for scalar perturbations ($s=0$ and $\xi=2$) the asymptotic behaviors of the wavefunction at the boundaries are
\beeq
\psi  \overset{r \rightarrow r_+}{\larrow}  (r-r_+)^{b \rho }~~\mbox{and}~~\psi \overset{r \rightarrow \infty}{\larrow}  r^{-2} ~,
\label{BC-scalar}
\eneq
where
\beeq
b= \frac{R^2 r_+}{3r_+^2+R^2}.
\label{b}
\eneq
We can now write the solution to the wave equation (\ref{eqWaveradial}) for scalar perturbations, with the desired behavior at the boundaries, in the form
\beeq
\psi(r) =  \left( \frac{r-r_+}{r} \right)^{b\rho}  \left(\frac{r_+}{r}\right)^{2} \sum_{n=0}^{\infty}a_n  \left( \frac{r-r_+}{r} \right)^{n
} .
\label{eq-WaveF-scalar}
\eneq
For electromagnetic ($s=1$) and gravitational ($s=2$) perturbations, where $\xi=0$, the asymptotic behaviors of the wavefunction at the boundaries are
\beeq
\psi  \overset{r \rightarrow r_+}{\larrow}  (r-r_+)^{b \rho }~~\mbox{and}~~\psi \overset{r \rightarrow \infty}{\larrow}  r^{-1} ~,
\label{}
\eneq
where the constant $b$ is given in Eq.\ (\ref{b}).
Therefore, for gravitational and electromagnetic perturbations, we write the solution to the wave equation (\ref{eqWaveradial}), with the desired behavior at the boundaries, in the form
\beeq
\psi(r) =  \left( \frac{r-r_+}{r} \right)^{b\rho}  \left(\frac{r_+}{r}\right) \sum_{n=0}^{\infty}a_n  \left( \frac{r-r_+}{r} \right)^{n
} .
\label{eq-WaveF-gravem}
\eneq

Substituting either (\ref{eq-WaveF-scalar}) or (\ref{eq-WaveF-gravem}) into the wave equation (\ref{eqWaveradial}) leads to a seven-term recurrence relation 
\beeq
\alpha_1 a_1 +\beta_1 a_0=0
\label{eq-rec1}
\eneq
\beeq
\alpha_2 a_2 +\beta_2 a_1+ \gamma_2 a_0=0
\label{}
\eneq
\beeq
\alpha_3 a_3 +\beta_3 a_2+ \gamma_3 a_1+ \delta_3 a_0=0
\label{}
\eneq
\beeq
\alpha_4 a_4 +\beta_4 a_3+ \gamma_4 a_2+ \delta_4 a_1 + \zeta_4 a_0=0
\label{}
\eneq
\beeq
\alpha_5 a_5 +\beta_5 a_4+ \gamma_5 a_3+ \delta_5 a_2 + \zeta_5 a_1 +\eta_5 a_0 =0
\label{}
\eneq
\beeq
\alpha_n a_n +\beta_n a_{n-1}+ \gamma_n a_{n-2}+ \delta_n a_{n-3} + \zeta_n a_{n-4} +\eta_n a_{n-5} +\kappa_n a_{n-6} =0
\label{eq-sevenrecurrence}
\eneq
where $n=6,7,8,\dots$ and $a_0$ is a constant which we can take to be 1.
The coefficients in the above recurrence relation for scalar perturbations are given in Appendix A. The coefficients for gravitational and electromagnetic perturbations are given in Appendix B.

The recurrence relation (\ref{eq-sevenrecurrence}) can be reduced to a six-term recurrence relation 
\beeq
\alpha_n' a_n +\beta_n' a_{n-1}+ \gamma_n' a_{n-2}+ \delta_n' a_{n-3} + \zeta_n' a_{n-4} +\eta_n' a_{n-5} =0
\eneq
by eliminating the $\kappa_n$ term using Gaussian elimination for $n \geq 6$ as follows:  
\begin{eqnarray}
\alpha_n' &=& \alpha_n \nonumber \\
\beta_n' &=& \beta_n - \frac{\alpha _{n-1}' \kappa_n}{\eta_{n-1}'}  \nonumber \\
\gamma_n' &=& \gamma_n - \frac{\beta_{n-1}' \kappa_n}{\eta_{n-1}'}   \nonumber \\
\delta_n' &=& \delta_n - \frac{\gamma_{n-1}' \kappa_n}{\eta_{n-1}'}   \nonumber \\
\zeta_n' &=& \zeta_n - \frac{\delta_{n-1}' \kappa_n}{\eta_{n-1}'}   \nonumber \\
\eta_n' &=& \eta_n - \frac{\zeta_{n-1}' \kappa_n}{\eta_{n-1}'}. \nonumber 
\end{eqnarray}
Repeating this procedure four times results in a three-term recurrence relation
\beeq
\alpha_n^{(4)} a_n +\beta_n^{(4)} a_{n-1}+ \gamma_n^{(4)} a_{n-2} =0.
\label{eq-threeterm}
\eneq

Following Leaver \cite{Leaver} we obtain a continued fraction
\begin{eqnarray}
\beta_1^{(4)} &=& \frac{\alpha_1^{(4)} \gamma_{2}^{(4)}}{\beta_{2}^{(4)}-\frac{\alpha_{2}^{(4)}\gamma_{3}^{(4)}}{\beta_{3}^{(4)} - \cdots}} \nonumber \\
&=& \frac{\alpha_1^{(4)}\gamma_{2}^{(4)}}{\beta_{2}^{(4)}-}\frac{\alpha_{2}^{(4)}\gamma_{3}^{(4)}}{\beta_{3}^{(4)} -}\frac{\alpha_{3}^{(4)}\gamma_{4}^{(4)}}{\beta_{4}^{(4)} -}\cdots 
\label{eqLeaver}
\end{eqnarray}
which is an equation in $\rho = -i\omega$. The solutions $\omega$ are the QNMs, with some exceptions that we discuss below.
To evaluate this numerically we must truncate the continued fraction at some depth. We use a technique developed by Nollert \cite{Nollert} to approximate the value of the tail end of the continued fraction.

In principle, we should be able to find all the QNMs by finding the roots to (\ref{eqLeaver}), but in practice our numerical methods only find a few roots using this single equation.  For this reason it is useful to ``invert" the equation.
The $n$-th inversion, for $n=2, 3, \dots$, of this continued fraction is defined to be \cite{Leaver} :
\beeq
\beta_n^{(4)} - \frac{\alpha_{n-1}^{(4)} \gamma_n^{(4)}}{\beta_{n-1}^{(4)} -} \frac{\alpha_{n-2}^{(4)} \gamma_{n-1}^{(4)}}{\beta_{n-2}^{(4)} -}\cdots \frac{\alpha_1^{(4)}\gamma_2^{(4)}}{\beta_1^{(4)}}
=
\frac{\alpha_{n}^{(4)} \gamma_{n+1}^{(4)}}{\beta_{n+1}^{(4)} -} \frac{\alpha_{n+1}^{(4)} \gamma_{n+2}^{(4)}}{\beta_{n+2}^{(4)} -} \frac{\alpha_{n+2}^{(4)}\gamma_{n+3}^{(4)}}{\beta_{n+3}^{(4)}-}\cdots
\label{eqInv}
\eneq
We will call (\ref{eqLeaver}) the $1$st inversion.  Different roots are easier to find numerically in different inversions.  Therefore, by changing inversions we are able to find as many roots as we want.  
However, inverting (\ref{eqLeaver}) introduces spurious roots (meaning they are not actual QNMs). 
What is common among these spurious roots  is that they satisfy the equation $\alpha_n^{(4)}\gamma_{n+1}^{(4)} = 0$.  One can show that if $\omega$ is a root that appears or disappears between the $n$th and $(n+1)$th inversion then it must be the case that $\alpha_n^{(4)}\gamma_{n+1}^{(4)} = 0$. Conversely, if $\omega$ satisfies $\alpha_n^{(4)}\gamma_{n+1}^{(4)} = 0$ then it will not be a root of any inversion past the $n$th (because the left-hand-side of Eq.\ (\ref{eqInv}), which is zero, will appear in the denominator in any subsequent inversion).  Thus spurious roots can only appear for one inversion.


Numerically, when solving for $\omega$, some roots will disappear between inversions simply because they have a smaller basin of attraction in the new inversion.  As far as we can tell, however, QNMs numerically appear across multiple inversions, whereas the roots of $\alpha_n^{(4)}\gamma_{n+1}^{(4)} = 0$
only appear for one inversion.  
We do not believe these roots are QNMs for the following reasons.
They do not appear in other numerical studies that use different methods, such as \cite{Horowitz, Konoplya0, Cardoso-L,Moss, Berti-K}, or in analytic studies, such as \cite{DG-HReal, EM-asymptoticQNM}.     
In fact, for each QNM found in previous numerical studies we find an infinite family of spurious roots branching from it.  
For the roots on these branches that we have checked, they all satisfy the equation  $\alpha_n^{(4)}\gamma_{n+1}^{(4)} = 0$. 
For the above reasons, we disregard any root that does not persist for at least two inversions.

In the particular case where  $\alpha_n^{(4)} = 0$, the recurrence relations do not determine unique values for the coefficients $a_i$, $i \ge n$.  
This contradicts the requirement that the solution to Eq.\ (\ref{eqWaveradial}) must be unique after imposing the boundary conditions.  Therefore, it appears that any $\omega$ that makes $\alpha_n^{(4)} = 0$ is not a QNM.  One specific example is the case of gravitational waves with $\ell=2$, $r_+ = 1$, where $\omega=-2 i$ is a root of the $1$st inversion of the continued fraction because, in this case, $\alpha_1^{(4)}=\beta_1^{(4)} = 0$ .  In fact, it has been proven by Miranda {\it et al.} \cite{Miranda}  that $-2i$ is {\it not} a QNM because the resulting wavefunction does not satisfy the boundary conditions.




\sxn{Numerical Process}
\vskip 0.3cm

For the purpose of brevity, in the rest of this paper, we choose units in which $R = 1$. That means $r$, $r_*$, $r_+$, $M$ and $t$ are expressed in units of $R$. The QNM frequency $\omega$ is in units of $R^{-1}$
and the units for the QNM potential $V (r)$ are $R^{-2}$.  All of our numerical calculations follow these units.

We can summarize our numerical process as follows. We begin by looking at computer generated complex plots of the lowest inversions near the origin. From this, we note the region containing the lowest roots. We then generate a grid of points in this region, which we use as initial guesses for a root finder to solve (\ref{eqInv}) over several inversions and two different depths.

Within each inversion, we keep the roots which exist in both depths (we consider roots to be the same if the distance between them is less than $10^{-8}$). From those roots, we then keep the ones which persist between inversions. What remains are roots clustered around the QNMs.  We throw out any clusters whose radius is greater than $10^{-8}$. We then identify the mathematical mean of each cluster, and keep it as a QNM.

With the first few QNMs in hand, we use the slope defined by the last two (most damped) QNMs to predict the location of the next QNM. We then use a root finder to locate that mode.  We repeat the process until the desired number of roots are found.

Finally, in the gravitational QNM spectrum, we observed some ``special" roots along the imaginary axis.  These roots are numerically very difficult to find.  Therefore, we also generate thousands of points along the imaginary axis, which we use as initial guesses for a root finder over several inversions and depths. We repeat the process explained above to ensure any roots found are QNMs.  
Purely imaginary QNMs appear to go to negative infinity as either $\ell$ increases or $r_+$ decreases.  Our method gets slower as the absolute value of $\omega$ increases because it requires a much higher depth in the continued fraction and thus higher precision in each calculation.  This makes them increasingly difficult to find.

\sxn{Results}
\vskip 0.3cm

Using the above techniques we have generated a large number of QNMs given in the tables below.  Tables I and II show the QNMs for gravitational and scalar perturbations, while tables III and IV show the QNMs for electromagnetic perturbations.

\vspace{0.5cm}
\footnotesize
\begin{tabular}{cccc}
	\multicolumn{4}{c} {Table I.  Gravitational QNMs for $\ell = 2$ with various $r_+$ values} \\
	\hline \hline
	$n$ & $r_+ = 0.4$ & $r_+ = 1$ & $r_+ = 50$ \\
	\hline
	& $	-8.6206055 i $ &  & $	-0.0266703 i $ \\
	$1$ & $	3.1620904-0.4309248 i $ &$	3.0331142-2.4042343 i $ &  $	92.5047566-133.1901656 i $ \\
	$2$ & $	4.4090815-1.5134459 i $ &$	4.9607292-4.8981936 i $ &$	158.1091839-245.8221619 i $ \\
	$3$ & $	5.8818960-2.6637516 i $ &  $6.9053582-7.2897272 i $ &  $	223.2778817-358.3814086 i $ \\
	$4$ & $	7.4543991-3.7971516 i $ &  $8.8546996-9.6604241 i $ &$	288.3378802-470.9112893 i $ \\
	$5$ & $	9.0724986-4.9039109 i $ & $	10.8078366-12.0234382 i $ & $	353.3567010-583.4291864 i $ \\
	$6$ & $	10.7109440-5.9893001 i $ & $	12.7638395-14.3826649 i $ & $	418.3561115-695.9413097 i $ \\
	$7$ & $	12.3584850-7.0599788 i $ & $	14.7219942-16.7396924 i $ & $	483.3450430-808.4502934 i $  \\
	$8$ & $	14.0102590-8.1209090 i $ &$	16.6817901-19.0953039 i $ &  $	548.3277639-920.9574123 i $ \\
	$9$ & $		15.6641429-9.1753252 i $ & $	18.6428630-21.4499373 i $ & $	613.3065437-1033.4633473 i $ \\
	$10$ & $17.3191697 - 10.2252823 i $ & $20.6049491 - 23.8038599 i$ & $	678.2826879 - 1145.9684908 i $ \\
	$25$ & $42.1793880 - 25.8188101 i $ & $50.0886404 - 59.0817570 i$ & $	1652.8356481 - 2833.5188349 i $ \\
	$50$ & $83.6542363 - 51.7010061 i $ & $99.2863968 - 117.8490937 i$ & $	3277.0366624 - 5646.0863655 i$ \\
	$100$ & $166.6415739 - 103.4216166 i $ & $197.7192949 - 235.3674961 i$ & $	6525.4254523 - 11271.2171329 i $ \\
	$250$ & $415.6592307 - 258.5370623 i$ & $493.0642611 - 587.9038157 i$ & $	16270.5850765 - 28146.6068515 i $ \\
	$500$ & $830.7198929 - 517.0407262 i$ & $985.3294717 - 1175.4550896 i$ & $	32512.5165965 - 56272.2556955 i $ \\
	$1000$ & $1660.8602827 - 1034.0357622 i $ & $1969.8735020 - 2350.5524365 i $ & $64996.3795527 - 112523.5531555 i $ \\
	\hline
\end{tabular}
\vspace{0.5cm}
\normalsize

\vspace{0.5cm}
\footnotesize
\begin{tabular}{cccc}
	\multicolumn{4}{c} {Table II.  Scalar QNMs for $\ell = 0$ with various $r_+$ values} \\
	\hline \hline
	$n$ & $r_+ = 0.4$ & $r_+ = 1$ & $r_+ = 50$ \\
	\hline
	$1$ & $	2.3629278-1.0064708 i $ & $	2.7982232-2.6712058 i $ & $	92.4936786-133.1932905 i $ \\
	$2$ & $	3.9785674-2.0727637 i $ & $	4.7584892-5.0375691 i $ & $	158.1007632-245.8243897 i $ \\
	$3$ & $	5.6170486-3.1195485 i $ & $	6.7192677-7.3944927 i $ & $	223.2708309-358.3832592 i $ \\
	$4$ & $	7.2637270-4.1610833 i $ & $	8.6822268-9.7485168 i $ & $	288.3316967-470.9129117 i $ \\
	$5$ & $	8.9144995-5.2002234 i $ & $	10.6466697-12.1012475 i $ & $	353.3511287-583.4306502 i $ \\
	$6$ & $	10.5676627-6.2380243 i $ & $	12.6121072-14.4532764 i $ & $	418.3509995-695.9426546 i $ \\
	$7$ & $	12.2223635-7.2749838 i $ & $	14.5782377-16.8048758 i $ & $	483.3402936-808.4515447 i $ \\
	$8$ & $	13.8781214-8.3113728 i $ & $	16.5448710-19.1561903 i $ & $	548.3233098-920.9585873 i $ \\
	$9$ & $	15.5346423-9.3473535 i $ & $	18.5118811-21.5073044 i $ & $	613.3023358-1033.4644586 i $ \\
	$10$ & $	17.1917350-10.3830299 i $ & $	20.4791821-23.8582713 i $ & $	678.2786896-1145.9695477 i $ \\
	$25$ & $	42.0743049-25.9043386 i $ & $	50.0028337-59.1163010 i $ & $	1652.8330984-2833.5195137 i $ \\
	$50$ & $	83.5707795-51.7596984 i $ & $	99.2235879-117.8736016 i $ & $	3277.0348548-5646.0868482 i $ \\
	$100$ & $	166.5781067-103.4627550 i $ & $	197.6738412-235.3848628 i $ & $	6525.4241725-11271.2174753 i $ \\
	$250$ & $	415.6166454-258.5630109 i $ & $	493.0349397-587.9148155 i $ & $	16270.5842665-28146.6070684 i $ \\
	$500$ & $	830.6889195-517.0590689 i $ & $	985.3085361-1175.4628718 i $ & $	32512.5160236-56272.2558490 i $ \\
	$1000$ & $	1660.8379527-1034.0487325 i $ & $	1969.8585979-2350.5579408 i $ & $	64996.3791475-112523.5532641 i $ \\
	\hline
\end{tabular}
\vspace{0.5cm}

\vspace{0.5cm}
\footnotesize
\begin{tabular}{cccc}
	\multicolumn{4}{c} {Table III.  Electromagnetic QNMs for $\ell = 1$ with various $r_+$ values} \\
	\hline \hline
	$n$ & $r_+ = 0.4$ & $r_+ = 1$ & $r_+ = 10$ \\
	\hline
	$1$  & $2.3298090-0.4411910 i $ & $	2.1630231-1.6990926 i $& $	-15.5057909 i $ \\
	& & & $	-28.4729072 i $ \\
	$2$& $	3.6810663-1.4970895 i $  & $	3.8438195-4.1519361 i $ &  $8.2213479 - 48.1622512 i $\\
	$3$ & $	5.1989859-2.5870273 i $ & $	5.6734729-6.5764558 i $ & $	20.6800750 - 71.4498997 i $ \\
	$4$ & $	6.7740513-3.6687077 i $ & $	7.5537242-8.9805378 i $ &$	33.2738238 - 94.3751851 i $ \\
	$5$ & $	8.3760254-4.7415670 i $ & $	9.4583847-11.3723844 i $ &  $	45.9659842 - 117.1903869 i $ \\
	$6$ & $	9.9929806-5.8077365 i $ & $	11.3772229-13.7563338 i $ &  $	58.7231435 - 139.9407128 i $ \\
	$7$ & $	11.6192611-6.8689450 i $ & $	13.3052644-16.1348222 i $ & $	71.5263567 - 162.6486670 i $ \\
	$8$ & $	13.2518153-7.9264047 i $ & $	15.2397363-18.5093289 i $ & $	84.3638491 - 185.3268708 i $ \\
	$9$  & $	14.8888361-8.9809611 i $ & $	17.1789429-20.8808126 i $& $	97.2278283 - 207.9830821 i$ \\
	$10$  & $	16.5291767-10.0332155 i $ & $	19.1217742-23.2499273 i $& $	110.1128787 - 230.6223968 i $ \\
	$25$ & $	41.2862642-25.6954654 i $ & $	48.4442954-58.6628581 i $ & $	304.5346705 - 569.3640074 i $ \\
	$50$  & $	82.7032719-51.6530851 i $ & $	97.5218549-117.5343929 i $& $	630.0334920 - 1132.9423059 i$ \\
	$100$  & $	165.6386458-103.4553474 i $ & $	195.8333278-235.1581149 i $ & $	1282.2614028 - 2259.3279839 i$ \\
	$250$  & $	414.5891139-258.6834199 i $& $	491.0151820-587.8349288 i $ & $	3241.0199331 - 5637.2325149 i $ \\
	$500$ & $	829.5981416-517.2744444 i $  & $	983.1552394-1175.4931925 i $& $6507.1087972 - 11266.1907295 i$ \\
	$998$  & $1656.3653866 - 1032.2898448 i $& $	1963.6351037 - 2345.9972772 i $ & $ 13014.3923657 - 22478.3521464 i $ \\
	\hline
\end{tabular}
\vspace{0.5cm}

\vspace{0.5cm}
\footnotesize
\begin{tabular}{ccc}
	\multicolumn{3}{c} {Table IV.  Electromagnetic QNMs for $\ell = 1$ with various $r_+$ values} \\
	\hline \hline
	$n$ & $r_+ = 20$ & $r_+ = 50$ \\
	\hline
	$1$  & $-30.2425917 i $& $	-75.0959881 i $ \\
	& $-59.1642407 i$  & $	-149.6543652 i $ \\
	$2$& $	-98.5793625 i $ &  $-226.9561884 i $\\
	& $	-102.4094897 i $ &   $	-291.7433132 i $ \\
	$3$ & $30.8947884-148.2650727 i $ & $	43.3602597-388.2720375 i $ \\
	$4$ &  $56.0241892-194.1418962 i $ &$	106.4001605-503.6209656 i $ \\
	$5$ &  $81.3308344-239.7811308 i $ &  $	169.7166198-617.7889596 i $ \\
	$6$ &  $	106.7526402-285.2820029 i $ &  $	233.2722272-731.5912326 i $ \\
	$7$ &  $132.2585800-330.6927893 i $ & $	297.0113691-845.1528498 i $ \\
	$8$ & $ 157.8283078-376.0405837 i $ & $	360.8940471-958.5463019 i $ \\
	$9$  &  $183.4479338-421.3419990 i $& $	424.8912893 - 1071.8162100 i$ \\
	$10$  &  $209.1076026-466.6079166 i $& $	488.9817705 - 1184.9918150 i $ \\
	$25$ &  $596.2180499-1143.8340830 i $ & $	1455.6342381 - 2877.9895678 i $ \\
	$50$  &  $1244.2718735-2270.5138181 i $& $	3073.7307600 - 5694.3937547 i$ \\
	$100$  &  $	2542.8189534-4522.3165071 i $ & $	6315.9664176 - 11323.2404591 i$ \\
	$250$  &  $	6442.5908521-11275.2135685 i $ & $	16052.9560069 - 28203.4504456 i $ \\
	$500$ &  $12945.1855388-22528.2790603 i $& $32288.6938102 - 56332.7117793 i$ \\
	$998$  &  $25900.8192803-44942.9434135 i $ & $	64636.4399310-112362.5921925 i $ \\
	\hline
\end{tabular}
\vspace{0.5cm}
\normalsize

In the tables, purely imaginary modes are numbered differently than the others.  In the gravitational case we do not assign these modes a number.  In the electromagnetic case, they appear in groups of two to which we assign the same number.  
This numbering will be explained in the next section where we discuss asymptotic behavior. 

\begin{figure}[th!]
	\begin{center}
		\includegraphics[height=7.5cm]{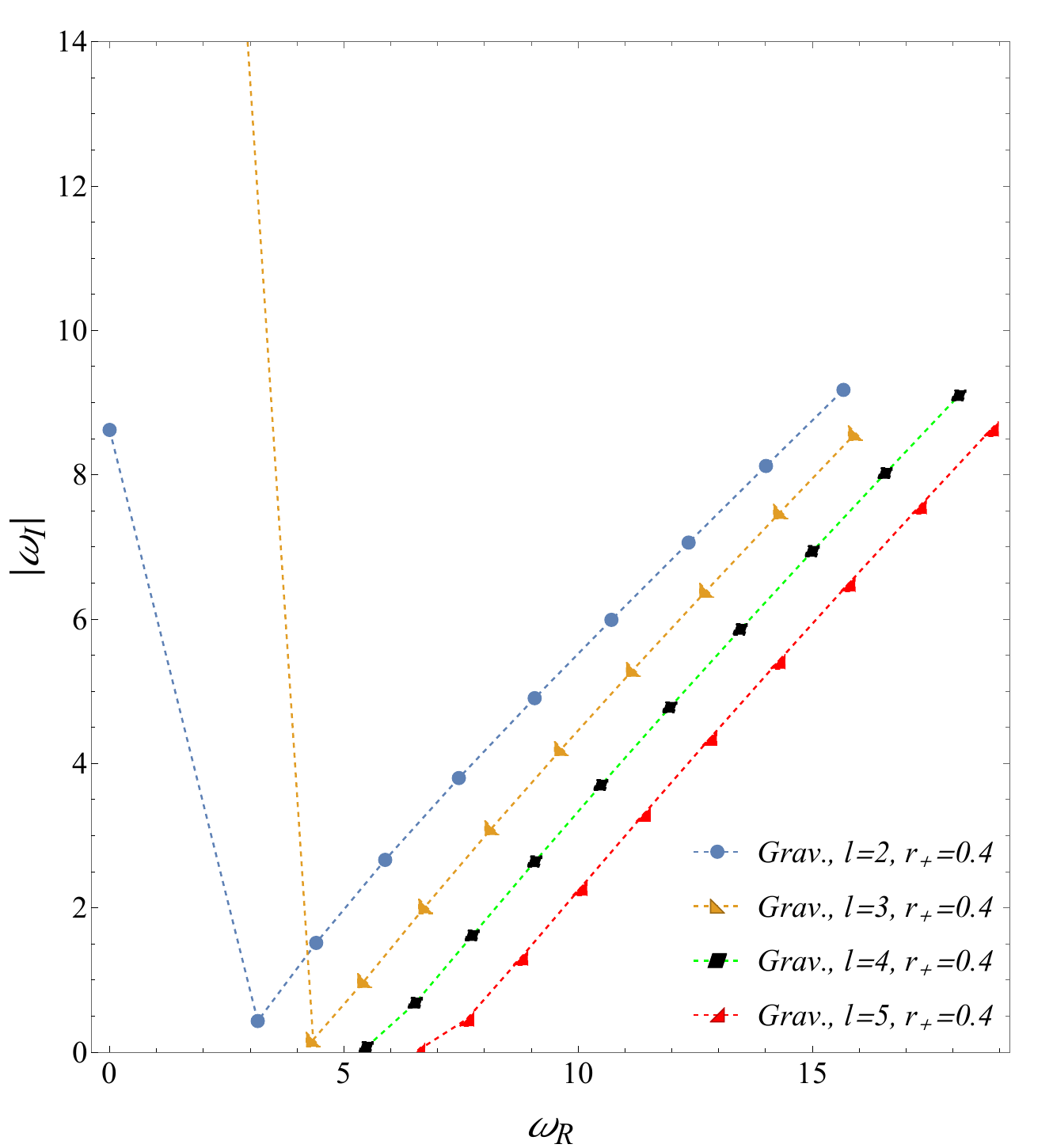}
		\includegraphics[height=7.8cm]{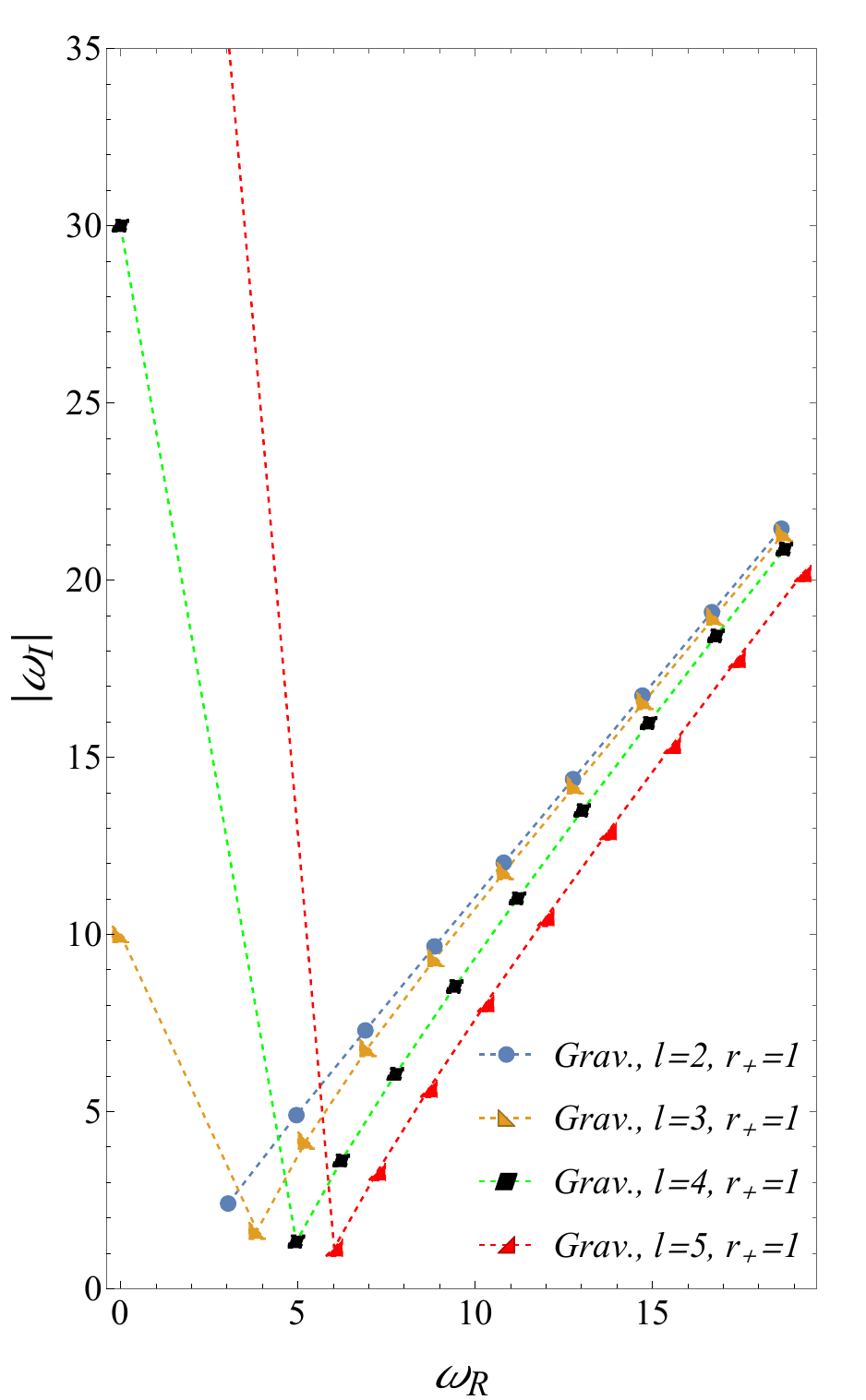}
		\includegraphics[height=6.3cm]{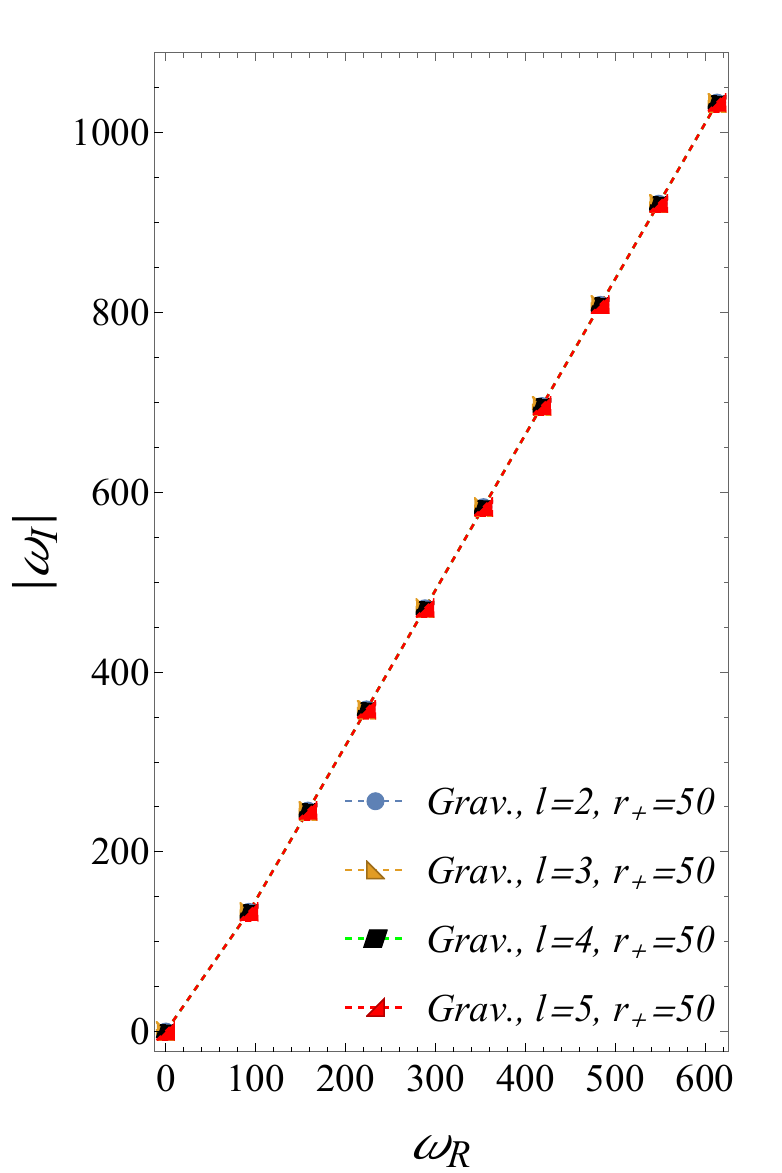}
	\end{center}
	\vspace{-0.7cm}
	\caption{\footnotesize QNM spectrum for gravitational odd parity perturbations with different values of $\ell$ and $r_+$.}
	\label{FigGravQNM}
\end{figure}

\begin{figure}[th!]
	\begin{center}
		\includegraphics[height=4.5cm]{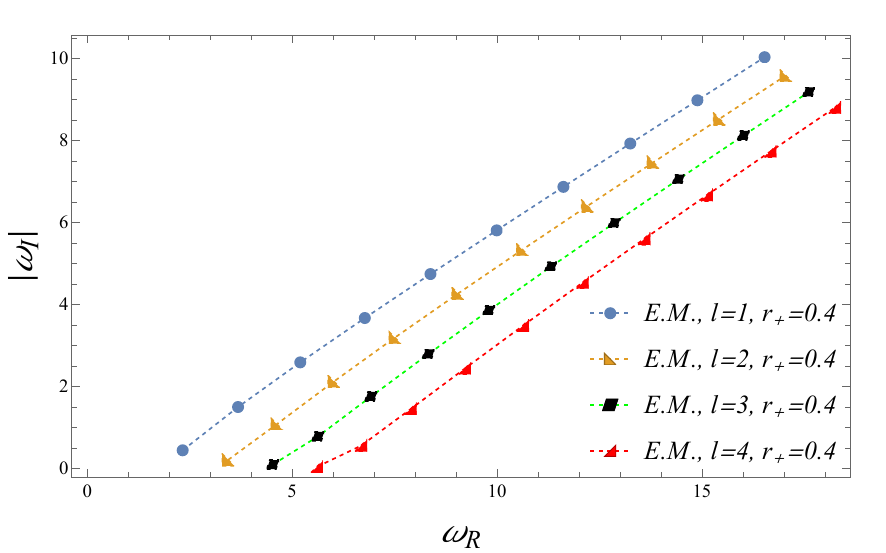}
		\includegraphics[height=5.6cm]{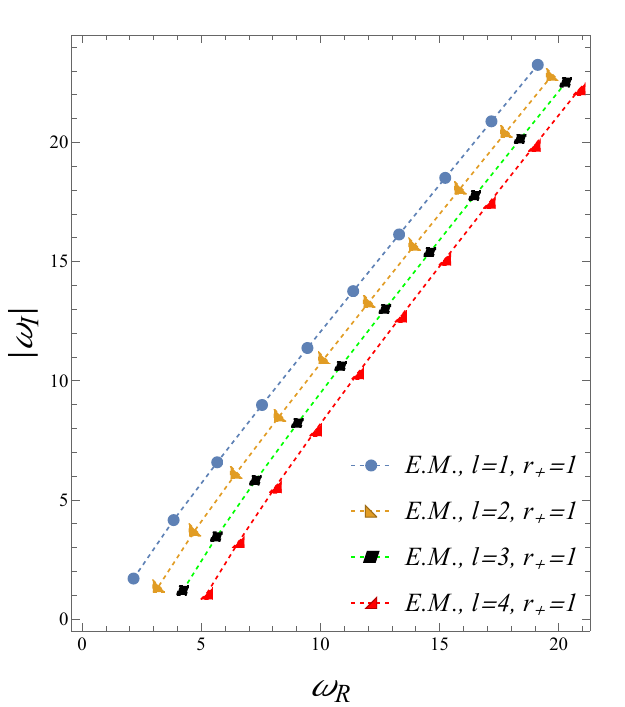}
		\includegraphics[height=6.2cm]{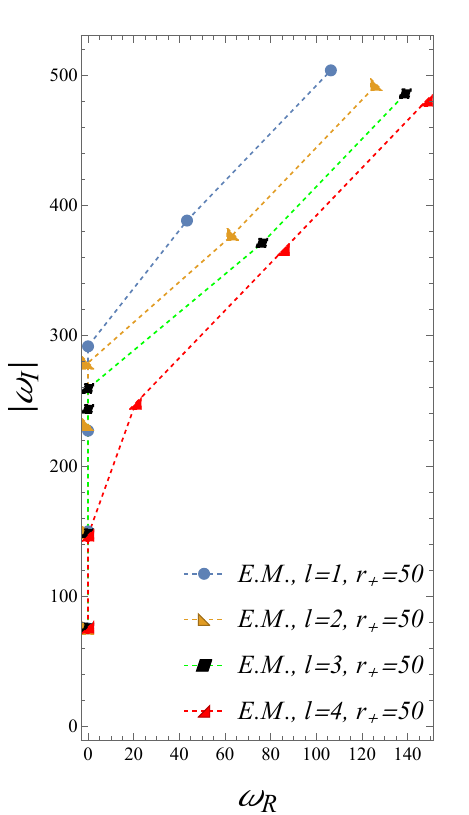}
	\end{center}
	\vspace{-0.7cm}
	\caption{\footnotesize QNM spectrum for electromagnetic perturbations with different values of $\ell$ and $r_+$. For $r_+ = 50$ the roots for $\ell=1, 2, 3, 4$ have bifurcated into four roots along the imaginary axis consistent with \cite{Wang, Fortuna}.}
	\label{FigEMQNM}
\end{figure}

\begin{figure}[th!]
	\begin{center}
		\includegraphics[width=7cm]{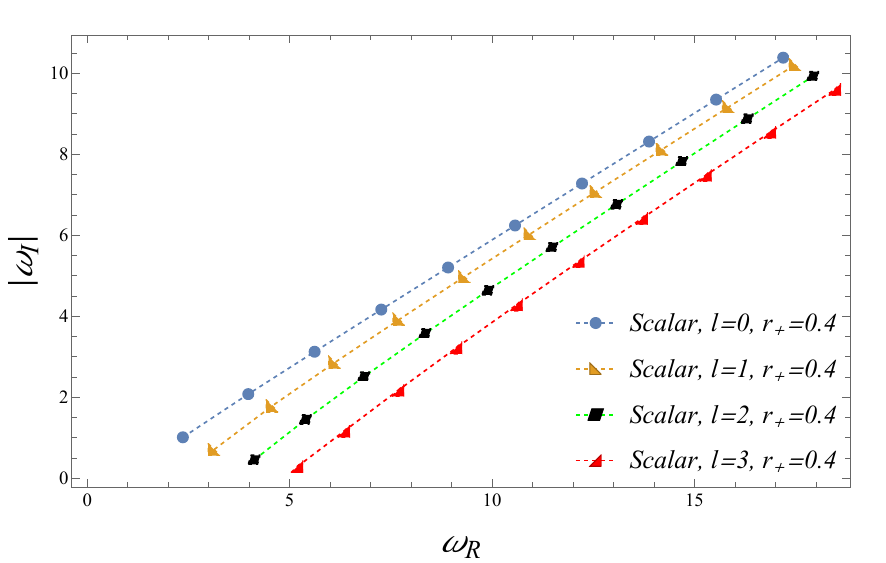}
		\includegraphics[width=4.8cm]{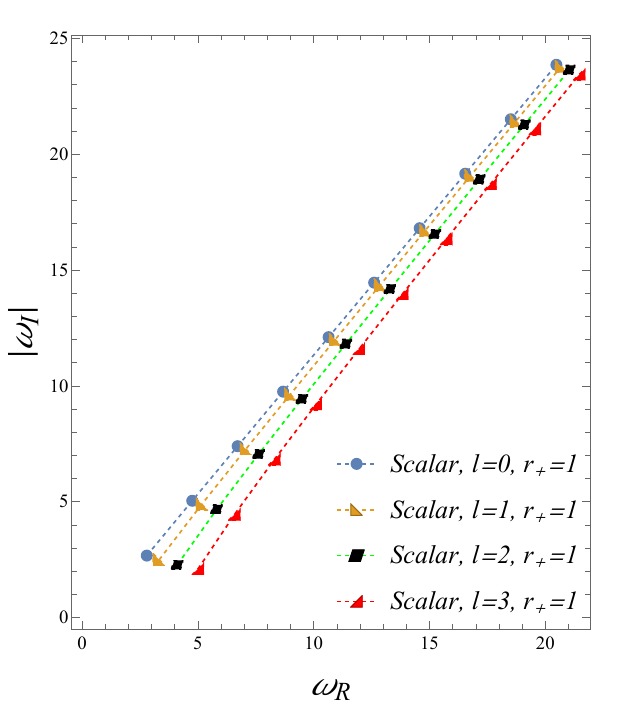}
		\includegraphics[width=4.1cm]{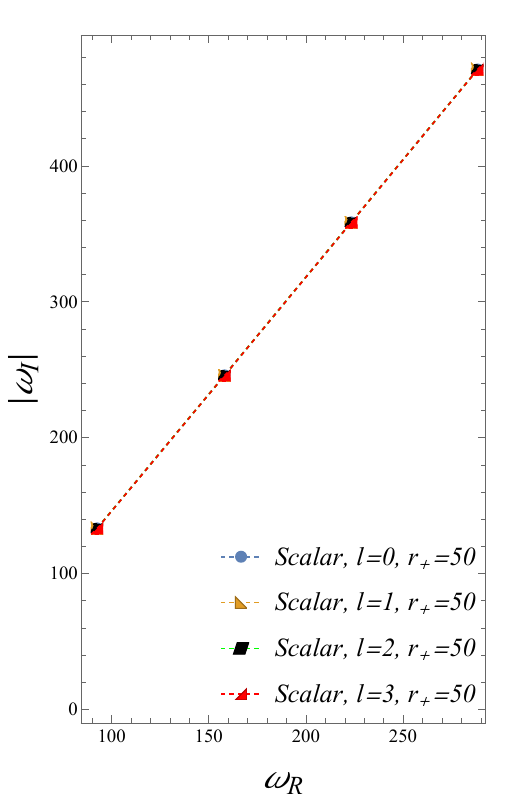}
	\end{center}  	
	\vspace{-0.7cm}
	\caption{\footnotesize QNM spectrum for scalar perturbations with different values of $\ell$ and $r_+$.}
	\label{FigScaQNM}
\end{figure}

We also determined the QNM spectrum for higher values of $\ell$.  Figures \ref{FigGravQNM}, \ref{FigEMQNM}, and \ref{FigScaQNM} show how these modes compare to the ones in the tables.
  
Where our results overlap with those previously published, we have agreement except for the items listed below:
\begin{itemize}
	\item For electromagnetic modes with $\ell = 2$ and $r_+ = 1$, the authors of \cite{Cardoso-L} find $2.496 - 1.579i$ as the first mode, where we find $3.223 - 1.384i$.   Our result agrees with \cite{Berti-K}. 
	\item   For electromagnetic modes with $\ell = 2$ and $r_+ = 50$, we find $-75.269i$ whereas \cite{Cardoso-L} finds $-75.139i$. Our result agrees with \cite{Berti-K}. 
	\item For gravitational modes with $\ell = 2$ and $r_+ = 1$, there is some confusion over whether $\omega = -2i$ is a QNM.   It first appears in \cite{Cardoso-L} with a question mark, it then appears in a graph in \cite{Moss}.  In \cite{Berti-K}, the authors say they find no evidence for it. Finally it appears again in \cite{Cardoso-K-L} (this time without the question mark).  
	As discussed at the end of section $2$, it is true that $–2i$ is a solution to the continued fraction equation.  However, this is a consequence of $\alpha_1^{(4)}=0$, which leads to a non-unique solution to the wave equation given by the recurrence relations (\ref{eq-rec1}-\ref{eq-sevenrecurrence}).  For that reason, we rejected it.  In fact, Miranda {\it et al.} \cite{Miranda} have proven that $-2i$ is {\it not} a QNM.
	\item For gravitational modes with $\ell = 2$, the authors of \cite{Berti-K} found $\omega = -1.335i/r_+$ is a good fit for the relation between $r_+$ and the purely imaginary $\omega$ for $r_+ \ge 10$.  This result was generalized to higher values of $\ell$ in \cite{Cardoso-K-L} and \cite{Miranda} to be $\omega=-i(l-1)(l+2)/(3r_+)$.  The behavior of the purely imaginary $\omega$ was previously studied in the context of AdS/CFT correspondence in \cite{Policastro} and \cite{Herzog}.
	For $r_+ \le 1$ we obtain a good fit with $\omega = 2.4416i - 4.4097i/r_+ $.  To get this fit we calculated the purely imaginary $\omega$ for $r_+ = 0.3, 0.4, ..., 1.0$.
	\item We do not find the ``highly real" QNMs that were calculated analytically in \cite{DG-HReal, D-HReal} for large black holes.  We searched in areas of the complex plane suggested by the asymptotic formula in \cite{DG-HReal}, but found no QNMs.
	However, there are issues that make searching for these modes troublesome.  Our search was limited to $\omega$ with a relatively small real part ($\omega_R < 65000, n\approx 500$) because larger values of $\omega$ take too long to compute.  The modes only are expected to appear for large $r_+$ which gives rise to QNMs with large real part. Our method gets slower as the absolute value of $\omega$ increases because it requires a much higher depth in the continued fraction and thus higher precision in each calculation. Considering the difficulty we had finding purely imaginary modes with high absolute value, it is possible the guesses need to be very close to the actual roots to find the highly real modes.
	We searched for these modes using $r_+ = 50$ for gravitational modes with $\ell = 2$ and scalar modes with $\ell = 0$.  We also searched for them by rewriting the QNM equation in the large $r_+$ limit (taking $f \approx -{2M \over r}+ \frac{r^2}{R^2}$).  In both cases we were unable to find the highly real QNMs.
\end{itemize}


\sxn{Comparison with the Asymptotic Spectrum}
\vspace{0.3cm}

For scalar and gravitational perturbations, the asymptotic (high overtone) values of $\omega$ have been calculated analytically in equation (3.39) of \cite{DG-HReal}  (and also equation (3.39) of \cite{Natario-S}) as\footnote{We have changed the indexing in \cite{DG-HReal}, which is completely arbitrary, to match that in our tables.}:



\beeq
\omega \eta = (n-1) \pi +\frac{5\pi}{4} + \frac{i}{2} \ln 2.
\label{eqAsymGrav}
\eneq
for $n=1, 2, \dots $ where $\eta$ is given by
\beeq
\eta = \sum_{j=0}^{2} \frac{1}{2 k_j}\ln\left(-\frac{1}{R_j}\right).
\eneq
$R_j$ are the zeros of the metric function $f(r)$ given in Eq.\ (\ref{function f}) and $k_j = f'(R_j)/2$.  

\begin{figure}[th!]
	\begin{center}
		\includegraphics[width=14.5cm]{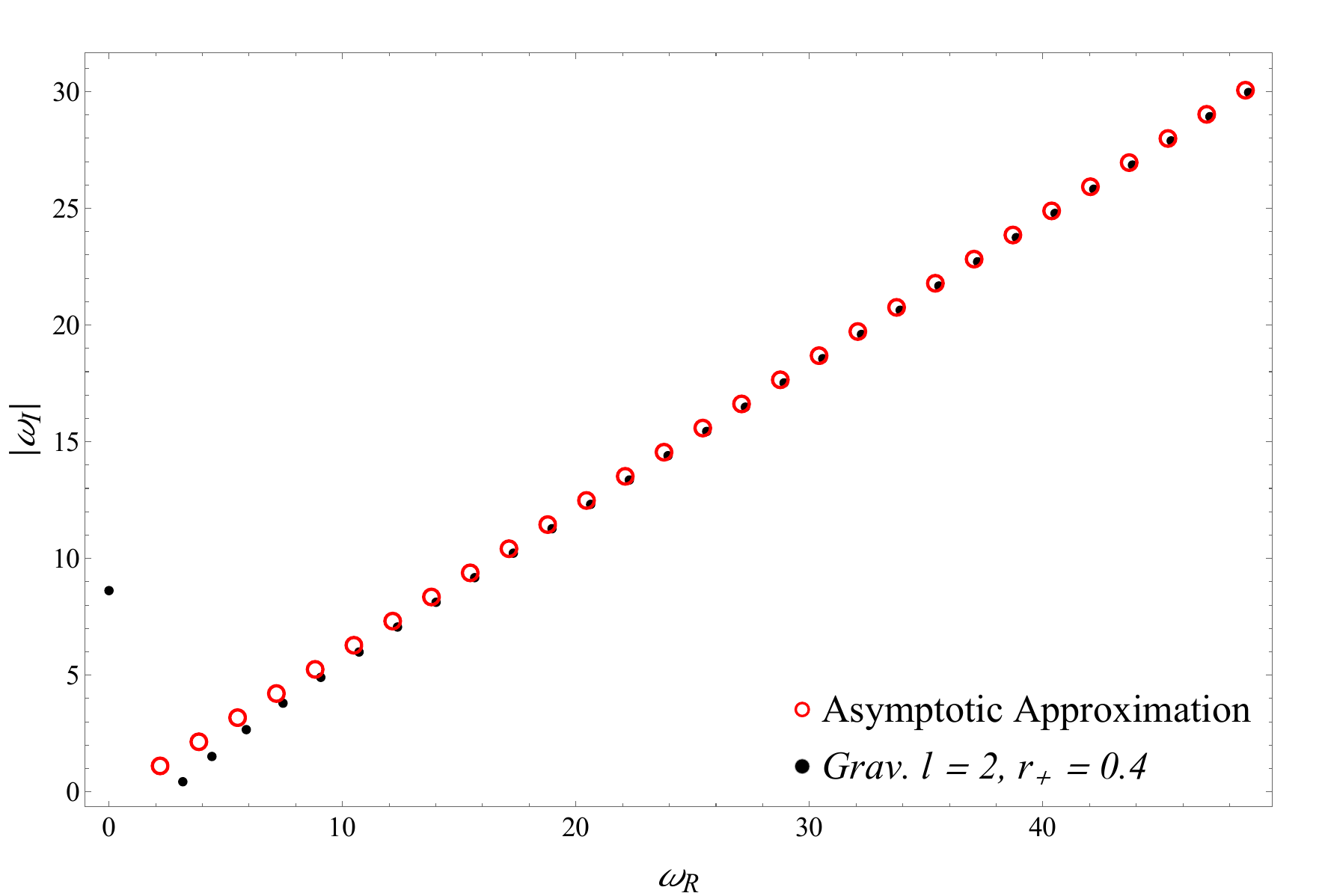}
	\end{center}
	\vspace{-0.7cm}
	\caption{\footnotesize Comparison of the asymptotic formula (\ref{eqAsymGrav}) with the actual QNMs.  The agreement is even better for larger $r_+$ values.}
	\label{FigAsymCompar}
\end{figure}

The value of $\eta$ and the explicit asymptotic formulas for $\omega$ for various $r_+$ values are given in Table V.  Although these formulas were analytically calculated as asymptotic approximations, the actual scalar and gravitational QNMs agree with the formulas remarkably well even for small values of $n$.  This agreement can be seen in Figure \ref{FigAsymCompar}.  There, we only show the gravitational modes for the $r_+ = 0.4$ case, but the agreement is similar for all $r_+$ values for both scalar and gravitational modes. 

The asymptotic values of $\omega$ for electromagnetic perturbations in the $4$-dimensional Schwarzschild-AdS spacetime can be derived from  Eq.\ (2.90) of \cite{EM-asymptoticQNM} by setting $j=1$ and $\beta=\pi/2$  (see Table 6 in \cite{EM-asymptoticQNM}).  This leads to the expression:
\beeq
\omega \eta = (n-1) \pi +\frac{5\pi}{4} -\frac{i}{2} \ln 3.
\label{eqEMWrongAsy}
\eneq
This gives the correct asymptotic behavior.  However, unlike the scalar and gravitational cases, this formula does not give good results for smaller $n$ values.  The approximation becomes worse for larger values of $r_+$. 
Based on our calculated values, we can modify Eq.\ (\ref{eqEMWrongAsy}) to  
\beeq
\omega \eta = (n-1) \pi +\frac{4\pi}{5} -i \ln [3.75(r_++0.3)],
\label{eqEMImprovedAsy}
\eneq
which gives better results for a wide range of $r_+$.  We show the difference in these two asymptotic formulas in figures \ref{FigAsymEM50} and \ref{FigAsymEMp4} below.

\begin{figure}[th!]
	\begin{center}
		\includegraphics[width=14.5cm]{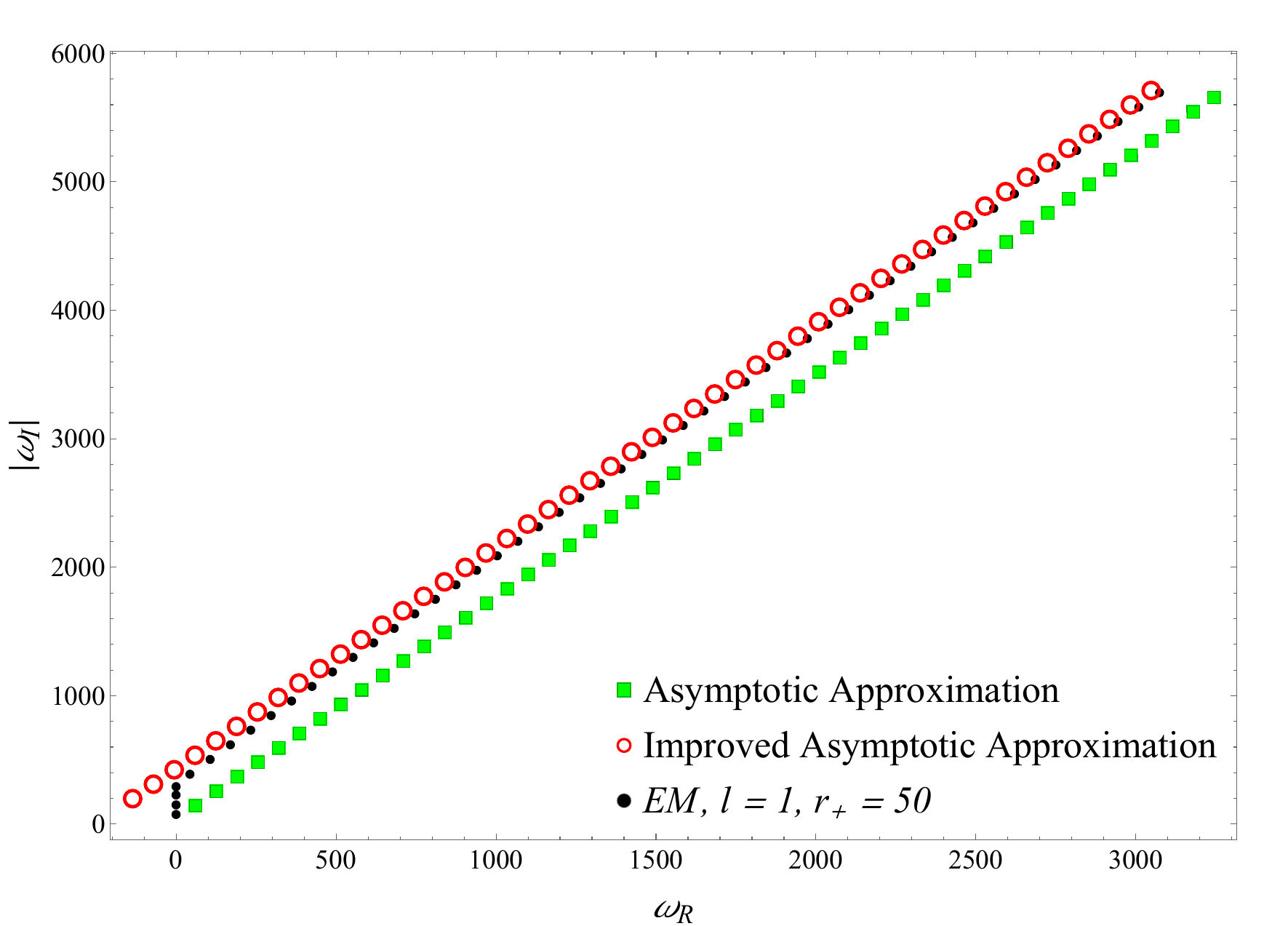}
	\end{center}
	\vspace{-0.7cm}
	\caption{\footnotesize Comparison of the asymptotic approximation (\ref{eqEMWrongAsy}) and the improved asymptotic approximation (\ref{eqEMImprovedAsy}) with the actual QNMs for $r_+ = 50$.}
	\label{FigAsymEM50}
\end{figure}
\begin{figure}[th!]
	\begin{center}
		\includegraphics[width=14.5cm]{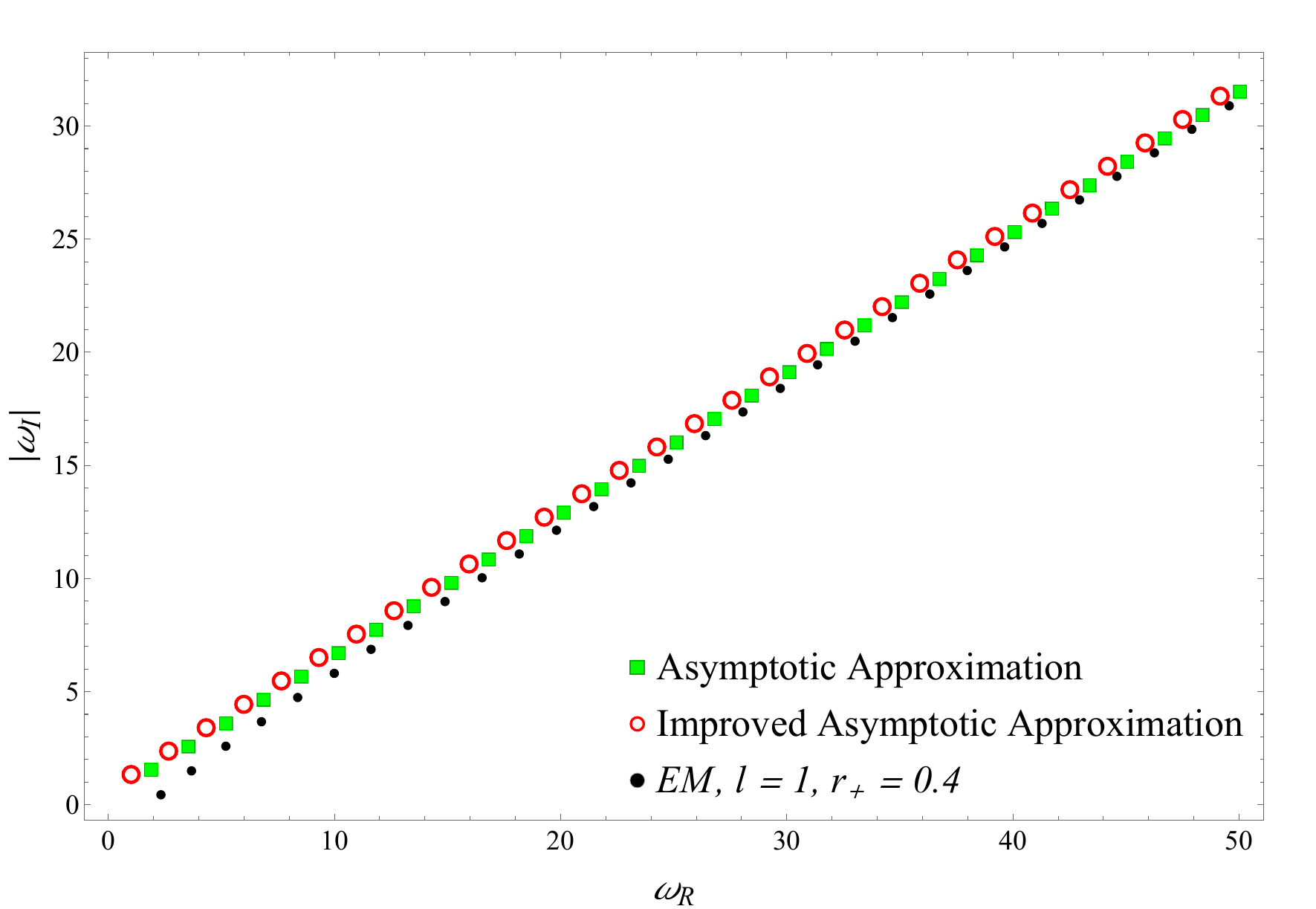}
	\end{center}
	\vspace{-0.7cm}
	\caption{\footnotesize Comparison of the asymptotic approximation (\ref{eqEMWrongAsy}) and the improved asymptotic approximation (\ref{eqEMImprovedAsy}) with the actual QNMs for $r_+ = 0.4$.}
	\label{FigAsymEMp4}
\end{figure}

To find the improved asymptotic formula (\ref{eqEMImprovedAsy}) we computed $\omega_n \eta - (n-1)\pi$ for $r_+ = 0.4, 1, 10, 20, 50$ and the largest value of $n$ we have in Tables III \& IV ($n = 998$).  We found the real part was approximately equal to $4\pi/5$ and we fit $\omega_I(r_+)$ with a function of the form $\ln[a(r_+ + b)]$.


\vspace{0.5cm}
\footnotesize
\begin{tabular}{cccc}
	\multicolumn{3}{l} {Table V.  Values of $\eta$ and the corresponding asymptotic formulas.}  \\
	\multicolumn{3}{l} {~~~~~~~~~~~~~Grav/Scalar, EM1, EM2 refer to Eqs.\ (\ref{eqAsymGrav}), (\ref{eqEMWrongAsy}), (\ref{eqEMImprovedAsy}), respectively.} \\
	\hline \hline
	$r_+$ &$\eta$& Asymptotic Formula for $\omega$ & Type \\
	\hline 
	$0.4$ & $1.3634 + 0.849079i$ & $(2.18944 - 1.10931 i) + (1.6603 - 1.03398 i) (n-1)$ & Grav/Scalar \\	
	&  & $(1.89459 - 1.58277 i) + (1.6603 - 1.03398 i) (n-1)$ & EM1\\	
	&  & $(1.01061 - 1.33722 i) + (1.6603 - 1.03398 i) (n-1)$ & EM2 \\
	\hline
	$1$ & $0.658045 + 0.785398 i$ &$(2.72065 - 2.72051 i) + (1.9691 - 2.35019 i) (n-1)$ & Grav/Scalar\\	
	&  & $(2.05045 - 3.28203 i) + (1.9691 -2.35019 i) (n-1)$ & EM1 \\	
	&  & $(0.390221 - 2.87305 i) + (1.9691 -2.35019 i) (n-1)$ & EM2\\	
	\hline
	$10$ & $0.0605905 + 0.104372 i$ &$(18.8202 - 26.6994 i) + (13.0693 - 22.5129 i) (n-1)$ & Grav/Scalar\\	
	&  & $(12.4003 - 30.4263 i) + (13.0693 - 22.5129 i) (n-1)$ & EM1\\
	&  & $(-15.7287 - 33.2109 i) + (13.0693 - 22.5129 i) (n-1)$ & EM2 \\		
	\hline
	$20$ & $0.0302464 + 0.0523163 i$ &$(37.4904 - 53.3876 i) + (26.0203 - 45.0065 i)(n-1)$ & Grav/Scalar \\	
	&  & $(40.3947 - 51.7085 i) + (26.0203 - 45.0065 i)  (n-1)$ & EM1 \\
	&  & $(-41.2494 - 71.8882 i) + (26.0203 - 45.0065 i)  (n-1)$ & EM2 \\								
	\hline
	$50$ & $ 0.012093 + 0.0209412i$ &$(93.6207 - 133.461 i) + (64.9677 - 112.503 i) (n-1)$ & Grav/Scalar \\	
	&  & $(61.5386 - 151.988 i) + (64.9677 - 112.503 i) (n-1)$ & EM1 \\
	&  & $(-135.665 - 198.36 i) + (64.9677 - 112.503 i) (n-1)$ & EM2 \\								
	\hline
\end{tabular}
\vspace{0.5cm}
\normalsize

Figures \ref{FigAsymCompar}, \ref{FigAsymEM50}, \ref{FigAsymEMp4} show that formulas  (\ref{eqAsymGrav}) and (\ref{eqEMImprovedAsy}) are good approximations to the QNM spectrum for all perturbations (gravitational, scalar, electromagnetic). The formulas work best for the lowest multipole number $\ell$.  However, for all values of $\ell$, the QNMs are well approximated by the values in the asymptotic formula for large enough $n$.

In the case of gravitational modes (Figure \ref{FigAsymCompar}), each QNM corresponds to exactly one mode given by Eq.\ (\ref{eqAsymGrav}) except for the purely imaginary one.  Therefore, the purely imaginary modes are ``special" in the sense that the asymptotic formula does not assign them a number. 
In Figure 5, we see that each value of $n$ in the formula for EM2 from Table V again corresponds to exactly one $\omega_n$ with the exception of the purely imaginary modes.  Each pair of purely imaginary modes corresponds to exactly one mode given by the asymptotic formula. It turns out that as $r_+$ increases, the modes move toward the imaginary axis.  When a mode hits the imaginary axis it splits into two purely imaginary modes.  This is the bifurcation discussed in \cite{Wang, Fortuna}.
This behavior was also noticed in an earlier work by Miranda {\it et al.} \cite{MirandaJHEP}.  A similar bifurcation was also reported in the electromagnetic QNM spectrum of rotating black strings in \cite{MorganJHEP}. 
Because of this bifurcation, in Tables III \& IV, the purely imaginary modes are grouped in pairs, corresponding to a single value of $n$. 
We find it intriguing that one could have predicted the bifurcation
by matching the QNMs to the asymptotic predictions.

\sxn{Summary}
\vskip .3cm

We conducted a comprehensive investigation of the gravitational, scalar, and electromagnetic QNM spectra of a Schwarzschild black hole in AdS spacetime using the numerical continued fraction method.  With a few noted exceptions, the low overtone QNMs that we found are consistent with previously obtained results in the literature that use other numerical techniques.   The intermediate and high overtone QNMs converge quickly to the asymptotic formulas previously obtained by analytic monodromy techniques.  Some of the highlights of our results are

\begin{itemize}
	\item   Using our numerical results, we were able to improve the asymptotic formula obtained in \cite{EM-asymptoticQNM} for high overtone electromagnetic QNMs.  Unlike the previous formula, our asymptotic formula works well for a wide range of QNMs and black hole sizes.   
	\item  The asymptotic formula obtained using analytic monodromy  techniques, in addition to matching well to the high overtone QNMs, is able to predict the bifurcation phenomenon at the low overtone region of the electromagnetic QNM spectrum.  
	\item In our search, we were unable to find  the ``highly real" QNMs that were calculated analytically in \cite{DG-HReal, D-HReal} for large black holes.  This makes it less likely that these modes exist.
We discussed issues that make searching for these modes troublesome. We consequently have no conclusive remark regarding the existence/nonexistence of these modes. 
	\item In the gravitational QNM spectrum, we observed some purely imaginary modes that appear to go to negative infinity as either $\ell$ increases or $r_+$ decreases. This is an interesting part of the QNM spectrum that can potentially be explored using the analytic monodromy techniques developed to calculate highly damped modes. We are surprised these analytic techniques have not previously revealed these modes. 
\end{itemize}

\vskip .6cm
\leftline{\bf Acknowledgments}
We are grateful to Kostas Kokkotas and Emanuele Berti for sharing with us their insight on the numerical method.  We also thank the anonymous referee for pointing out a crucial piece of information relating to the purely imaginary modes, which led us to correct some of our results. 

\vskip .6cm
\leftline{\bf \large Appendix A Scalar Coefficients}
\label{appendix:scalar}
\vskip .3cm

For scalar perturbations, the coefficients for the recurrence relations (\ref{eq-rec1} - \ref{eq-sevenrecurrence}) are
\begin{eqnarray}
\alpha_n &=& -(1+3r_+^2)^2(n+b\rho)^2+r_+^2 \rho^2 .\nonumber
\label{alpha}
\end{eqnarray}
\begin{eqnarray}
	\beta_n &=& 3+\ell (\ell + 1)(1+3r_+^2)-6b\rho +6b^2\rho^2  \nonumber \\
	&&+15r_+^2 +18r_+^4 +6n^2 (1+5r_+^2+6 r_+^4)-27 r_+^2 b \rho-27 r_+^4 b \rho-2 r_+^2  \rho^2 \nonumber \\
	&& +30 r_+^2 b^2\rho^2+36 r_+^4 b^2 \rho^2-3n(1+3r_+^2)(2+3r_+^2-4b\rho-8r_+^2 b \rho). \nonumber
	\label{beta}
\end{eqnarray}
\begin{eqnarray}
	\gamma_n &=& 2 r_+^2 (1+n+b\rho)-6+5\ell (\ell + 1) +17n -13n^2+17 b \rho-26nb \rho-13b^2 \rho^2+r_+^2 \rho^2 \nonumber \\
	&&-  (1+r_+^2)^2 \left[64+60 n^2 -107 b \rho +60 b^2 \rho^2 -107 n+120n b \rho \right]  \nonumber \\
	&&-  2 r_+^2 (1+r_+^2) (1+4n+4 b \rho) \nonumber \\
	&&-  (1+r_+^2) \left[-49+9\ell (\ell + 1)+94n-58n^2+94b \rho-116n b \rho -58 b^2 \rho^2  \right].\nonumber
	\label{gamma}
\end{eqnarray}
\begin{eqnarray}
	\delta_n &=&  -4\ell (\ell + 1)+6(-1+n+b \rho)^2 + 2 r_+^2(1+r_+^2) (n+b \rho)\nonumber \\
	&& + (1+r_+^2)^2 \left[ 117+54 n^2-146 b \rho +54b^2 \rho^2+2n(-73+54 b \rho)  \right]\nonumber \\
	&& + (1+r_+^2) \left[-69+10\ell (\ell + 1)+98 n-40 n^2+98 b \rho-80 n b \rho-40 b^2 \rho^2\right] .\nonumber 
	\label{delta}
\end{eqnarray}
\begin{eqnarray}
	\zeta_n &=& -2 + \ell (\ell + 1)+3n -n^2+3b \rho -2 n b \rho -b^2 \rho^2   \nonumber \\
	&&+ (1+r_+^2) \left[ -43+5 \ell (\ell + 1)+48 n -14 n^2+48 b \rho -28 n b \rho -14 b^2 \rho^2 \right]
	\nonumber \\
	&& -(1+r_+^2)^2 \left[107 +28 n^2 -105 b \rho +28 b^2 \rho^2 +7 n(8 b \rho-15) \right] .\nonumber 
	\label{zeta}
\end{eqnarray}
\begin{eqnarray}
	\eta_n &=&  (1+r_+^2) \left[-10+\ell (\ell + 1)+9n-2n^2+9 b \rho -4n b \rho -2 b^2 \rho^2  \right] \nonumber \\
	&& + (1+r_+^2)^2  \left[ 49 + 8n^2-39 b \rho +8 b^2 \rho^2 +n(16 b \rho-39) \right] .\nonumber
	\label{eta}
\end{eqnarray}
\begin{eqnarray}
	\kappa_n &=&  - (1+r_+^2)^2  (-3+n+b \rho)^2.\nonumber
	\label{kappa}
\end{eqnarray}

\vspace{.5cm}
\leftline{\bf \large Appendix B Gravitational and Electromagnetic Coefficients}
\vspace{.5cm}

For electromagnetic and gravitational  perturbations, the coefficients for the recurrence relations (\ref{eq-rec1} - \ref{eq-sevenrecurrence}) are
\begin{eqnarray}
	\alpha_n &=& - (1+3r_+^2)^2 (n+b\rho)^2+r_+^2 \rho^2. \nonumber
	\label{alpha}
\end{eqnarray}
\begin{eqnarray}
	\beta_n &=& 4+\ell (\ell + 1)(1+3r_+^2)-8b\rho +6b^2\rho^2 -s^2(1+4 r_+^2+3 r_+^4) \nonumber \\
	&&+19r_+^2 +21r_+^4 +6n^2 (1+5r_+^2+6 r_+^4)-39 r_+^2 b \rho-45 r_+^4 b \rho-2 r_+^2  \rho^2 \nonumber \\
	&& +30 r_+^2 b^2\rho^2+36 r_+^4 b^2 \rho^2-n(1+3r_+^2)(8+15r_+^2-12b\rho-24r_+^2 b \rho).\nonumber
	\label{beta}
\end{eqnarray}
\begin{eqnarray}
	\gamma_n &=&-22+5\ell (\ell + 1) -13n^2-2 r_+^2 +33 b \rho+2r_+^2 b \rho \nonumber \\
	&&~~~~~-13 b^2 \rho^2+r_+^2 \rho^2+n(33+2r_+^2-26 b \rho) \nonumber \\
	 &&- (1+r_+^2)^2 \left[124+60 n^2 -12 s^2 -161 b \rho +60 b^2 \rho^2 -161 n+120n b \rho \right]  \nonumber \\
	&&- (1+r_+^2) \Big[-114+9\ell (\ell + 1)-58n^2-10r_+^2+7s^2+154 b \rho +8 r_+^2 b \rho \nonumber \\
	&&~~~~~~~~~~~~~~~~~~ -58 b^2 \rho^2 +2n(77+4r_+^2-58 b \rho) \Big] .\nonumber
	\label{gamma}
\end{eqnarray}
\begin{eqnarray}
	\delta_n &=&  20 -4\ell (\ell + 1)+6n^2-22 b \rho +6 b^2 \rho^2 -22n+12nb\rho  \nonumber \\
	&& +(1+r_+^2)^2 \left[ 224+54 n^2-19s^2-212 b \rho +54b^2 \rho^2+4n(-53+27 b \rho)  \right]\nonumber \\
	&& + (1+r_+^2) \Big[-156+10\ell (\ell + 1)-40 n^2-4r_+^2 +9s^2+154 b \rho+2 r_+^2 b \rho \nonumber \\ 
	&& ~~~~~~~~~~~~~~~~~~-40  b^2 \rho^2+2n (77+r_+^2-40 b \rho) \Big] .\nonumber
	\label{delta}
\end{eqnarray}
\begin{eqnarray}
	\zeta_n &=& -6 + \ell (\ell + 1)+5n -n^2+5b \rho -2 n b \rho -b^2 \rho^2 
 \nonumber \\
	&& 	-(1+r_+^2) \left[ -94+5 \ell (\ell + 1)+72 n -14 n^2+5s^2+72 b \rho -28 n b \rho -14 b^2 \rho^2 \right]\nonumber \\
	&& - (1+r_+^2)^2 \left[200 +28 n^2 -15s^2-147 b \rho +28 b^2 \rho^2 +7 n(8 b \rho-21) \right] .\nonumber 
	\label{zeta}
\end{eqnarray}
\begin{eqnarray}
	\eta_n &=&  (1+r_+^2) \left[-21+\ell (\ell + 1)+13n-2n^2+s^2 +13 b \rho -4n b \rho -2 b^2 \rho^2  \right] \nonumber \\
	&& + (1+r_+^2)^2  \left[ 89 + 8n^2-6 s^2-53 b \rho +8 b^2 \rho^2 +n(16 b \rho-53) \right] .\nonumber
	\label{eta}
\end{eqnarray}
\begin{eqnarray}
	\kappa_n &=&  - (1+r_+^2)^2  [n^2-s^2 -2n(4-b\rho)+(4-b\rho)^2].\nonumber
	\label{kappa}
\end{eqnarray}

\newpage


\def\jnl#1#2#3#4{{#1}{\bf #2} (#4) #3}

\def\Zphys{{\em Z.\ Phys.} }
\def\jssc{{\em J.\ Solid State Chem.} }
\def\jpsJ{{\em J.\ Phys.\ Soc.\ Japan} }
\def\ptps{{\em Prog.\ Theoret.\ Phys.\ Suppl.\ } }
\def\PTP{{\em Prog.\ Theoret.\ Phys.\  }}
\def\LNC{{\em Lett.\ Nuovo.\ Cim.\  }}
\def\LRR{{\em Living \ Rev.\ Relative.} }
\def\JMP{{\em J. Math.\ Phys.} }
\def\NPB{{\em Nucl.\ Phys.} B}
\def\NP{{\em Nucl.\ Phys.} }
\def\PLB{{\em Phys.\ Lett.} B}
\def\PL{{\em Phys.\ Lett.} }
\def\PRL{\em Phys.\ Rev.\ Lett. }
\def\PRB{{\em Phys.\ Rev.} B}
\def\PRD{{\em Phys.\ Rev.} D}
\def\PRX{{\em Phys.\ Rev.} X~}
\def\PR{{\em Phys.\ Rev.} }
\def\PRe{{\em Phys.\ Rep.} }
\def\AP{{\em Ann.\ Phys.\ (N.Y.)} }
\def\RMP{{\em Rev.\ Mod.\ Phys.} }
\def\ZPC{{\em Z.\ Phys.} C}
\def\SCI{\em Science}
\def\CMP{\em Comm.\ Math.\ Phys. }
\def\MPLA{{\em Mod.\ Phys.\ Lett.} A}
\def\IJMPB{{\em Int.\ J.\ Mod.\ Phys.} B}
\def\cmp{{\em Com.\ Math.\ Phys.}}
\def\JPA{{\em J.\  Phys.} A}
\def\CQG{\em Class.\ Quant.\ Grav.~}
\def\ATMP{\em Adv.\ Theoret.\ Math.\ Phys.~}
\def\PRSA{{\em Proc.\ Roy.\ Soc.\ Lond.} A }
\def\IJTP{\em Int.\ J.\ Theor.\ Phys.~}
\def\GERG{\em Gen.\ Rel.\ Grav.~}
\def\JHEP{\em JHEP~}
\def\ibid{{\em ibid.} }

\vskip 1cm

\leftline{\bf References}

\renewenvironment{thebibliography}[1]
        {\begin{list}{[$\,$\arabic{enumi}$\,$]}  
        {\usecounter{enumi}\setlength{\parsep}{0pt}
         \setlength{\itemsep}{0pt}  \renewcommand{\baselinestretch}{1.2}
         \settowidth
        {\labelwidth}{#1 ~ ~}\sloppy}}{\end{list}}


\end{document}